\documentclass[aps,prb,twocolumn,notitlepage,citeautoscript,superscriptaddress,nopacs,10pt]{revtex4-1}

\usepackage{times}
\usepackage{amsmath,amssymb,mathrsfs,bm,setspace}
\usepackage{dcolumn}
\usepackage{graphicx}
\usepackage{latexsym}
\usepackage{hyperref}
\usepackage{comment}

\usepackage{bm}
\usepackage{epsfig}
\usepackage{amsmath}
\usepackage{color}
\usepackage{ulem}

\usepackage{xparse}
\usepackage{soul}
\usepackage{xcolor}

\begin{document}

\title{
Theoretical evidence of spin-orbital-entangled $J_{\mathbf{eff}}$=1/2 state in the 3$d$ transition metal oxide CuAl$_2$O$_4$
}

\author{Choong~H. \surname{Kim}}
\email[]{chkim82@snu.ac.kr}
\affiliation{Center for Correlated Electron Systems, Institute for Basic Science (IBS), Seoul 08826, Republic of Korea}
\affiliation{Department of Physics and Astronomy, Seoul National University, Seoul 08826, Republic of Korea}

\author{Santu \surname{Baidya}}
\affiliation{Center for Correlated Electron Systems, Institute for Basic Science (IBS), Seoul 08826, Republic of Korea}
\affiliation{Department of Physics and Astronomy, Seoul National University, Seoul 08826, Republic of Korea}

\author{Hwanbeom \surname{Cho}}
\affiliation{Center for Correlated Electron Systems, Institute for Basic Science (IBS), Seoul 08826, Republic of Korea}
\affiliation{Department of Physics and Astronomy, Seoul National University, Seoul 08826, Republic of Korea}

\author{Vladimir~V. \surname{Gapontsev}}

\affiliation{M.N. Miheev Institute of Metal Physics of Ural Branch of Russian Academy of Sciences, 620137 Ekaterinburg, Russia}

\author{Sergey~V. \surname{Streltsov}}

\affiliation{M.N. Miheev Institute of Metal Physics of Ural Branch of Russian Academy of Sciences, 620137 Ekaterinburg, Russia}
\affiliation{Ural Federal University, 620002 Ekaterinburg, Russia}

\author{Daniel~I. \surname{Khomskii}}

\affiliation{II. Physikalisches Institut, Universit\"{a}t zu K\"{o}ln, D-50937 K\"{o}ln, Germany}

\author{Je-Geun \surname{Park}}
\affiliation{Center for Correlated Electron Systems, Institute for Basic Science (IBS), Seoul 08826, Republic of Korea}
\affiliation{Department of Physics and Astronomy, Seoul National University, Seoul 08826, Republic of Korea}

\author{Ara \surname{Go}}
\email[]{arago@ibs.re.kr}

\affiliation{Center for Theoretical Physics of Complex Systems, Institute for Basic Science (IBS), Daejeon 34126, Republic of Korea}

\author{Hosub \surname{Jin}}
\email[]{hsjin@unist.ac.kr}
\affiliation{Department of Physics, Ulsan National Institute of Science and Technology (UNIST), Ulsan 44919, Republic of Korea}

\begin{abstract}
The spin-orbital-entangled Kramers doublet, known as the $J_{\mathbf{eff}}$=1/2 pseudospin driven by large spin-orbit coupling (SOC),
appears in layered iridates and $\alpha$-RuCl$_3$, manifesting a relativistic Mott insulating phase. Such entanglement, however, seems barely
attainable in 3$d$ transition metal oxides, where the SOC is small and the orbital angular momentum is easily quenched.
Based on the density functional theory calculations, we report the CuAl$_2$O$_4$ spinel as the possible example of a $J_{\mathbf{eff}}$=1/2
Mott insulator in 3$d$ transition metal compounds. With the help of strong electron correlations, the $J_{\mathbf{eff}}$=1/2 state
can survive the competition with an orbital-momentum-quenched $S$=1/2 state in the $d$$^9$ configuration of CuO$_4$ tetrahedron.
From the dynamical mean field theory calculations, the electron-addition spectra probing unoccupied states are
well described by the $j_{\mathbf{eff}}$=1/2 hole state, whereas electron-removal spectra have a rich multiplet structure.
The fully relativistic entity found in CuAl$_2$O$_4$ provides new insight into the untapped regime where the spin-orbital-entangled Kramers pair
coexists with strong electron correlation.
\end{abstract}

\maketitle

Transition metal oxides exhibit various competing phases and exotic phenomena depending on how they react to the rich degeneracy of the
$d$-orbital~\cite{Imada1998,Kugel1982,Khomskii2014}. Spin-orbit coupling (SOC) reduces this degeneracy in a unique way by providing
a spin-orbital-entangled ground state. In particular, the spin-orbital-entangled $J_{\mathbf{eff}}$=1/2 Kramers doublet has emerged
in the 4$d$ and 5$d$ transition metal compounds with the $t_{\rm 2g}^5$ configuration
 due to a large atomic spin-orbit coupling (SOC) assisted by moderate electron correlation\cite{Kim2008,Kim2009c,Plumb2014,Kim2014,Jackeli2009}.
A variety of novel phenomena has also risen from the $J_{\mathbf{eff}}$=1/2 state,
including a 5$d$ analogue to high $T$$_{\rm c}$ cuprate in a square lattice~\cite{Wang2011d,Kim2014d}, topological insulators~\cite{Shitade2009,Kim2012},
the Kitaev model~\cite{Jackeli2009,Chaloupka2010,Kitagawa2018,Plumb2014,Winter2017}, Weyl semi-metals~\cite{Wan2011}, axion insulators~\cite{Go2012},
and so on~\cite{Rau2016}. It is interesting to ask how the spin-orbital-entangled state behaves under strong electron correlation~\cite{Witczak-Krempa2013b}.
However, this question remains hypothetical, simply because no transition metals can possibly possess both large SOC and strong electron correlation
simultaneously. If we take large SOC strength as a prerequisite for the spin-orbital entanglement in the $t_{\rm 2g}^5$ configuration~\cite{Martins2016},
the intriguing strongly correlated $J_{\rm eff}$=1/2 state in real materials seems impractical. The Co$^{2+}$ environment has been suggested as a promising
candidate for the strongly correlated spin-orbital-entangled state ~\cite{Liu2018,Sano2018}, but it is yet to be confirmed.

\begin{figure}
  \centering
  \includegraphics[width=0.45\textwidth,type=pdf,ext=.pdf,read=.pdf]{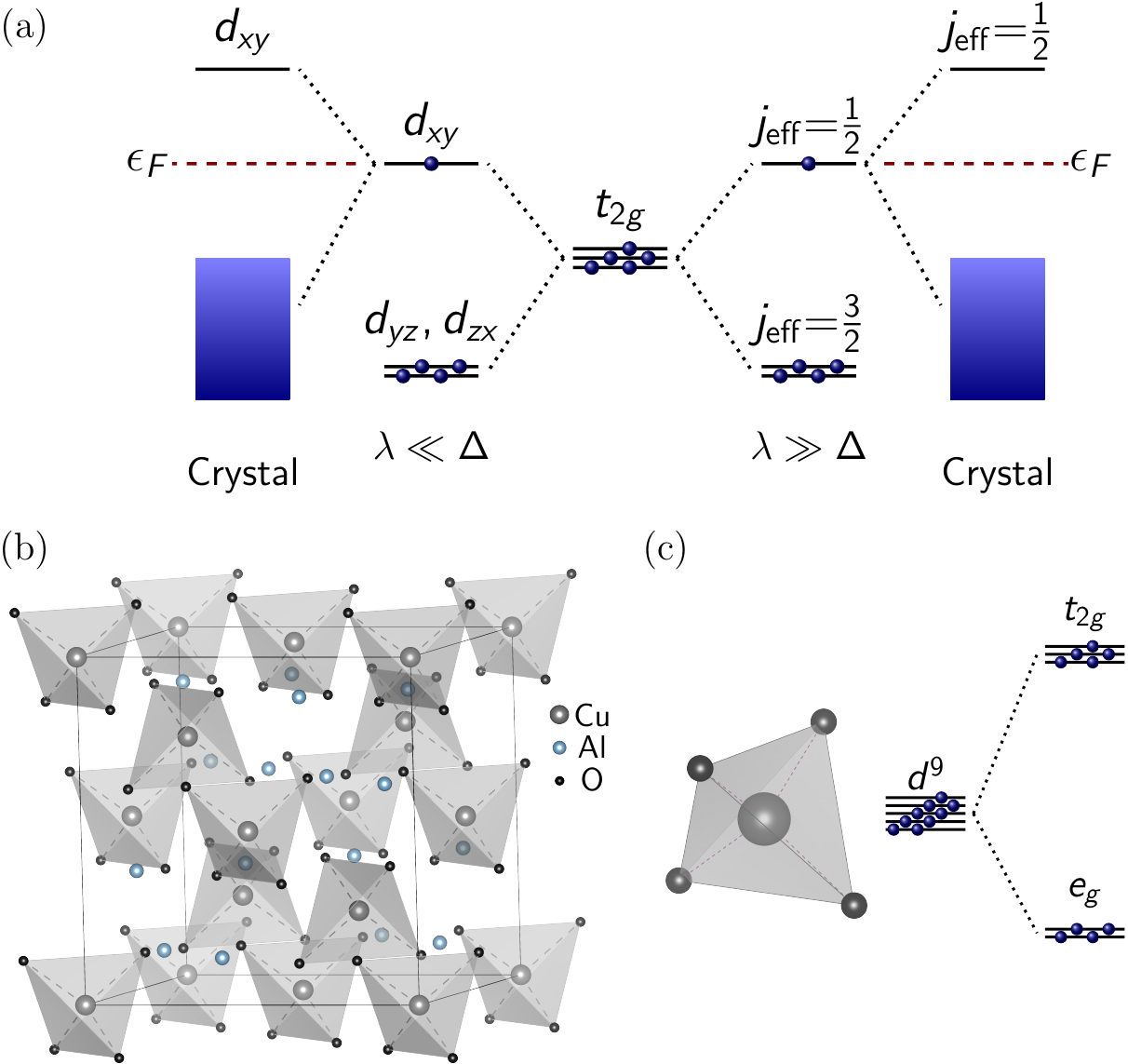}
  \caption{
    (a) Two possible ground states from the competition between Jahn-Teller distortion ($\Delta$) and spin orbit coupling ($\lambda$),
    resulting in $J_{\rm eff}$=1/2 and $S$=1/2 states, respectively.
    (b) The crystal structure of CuAl$_2$O$_4$. The grey, light blue, and black spheres represent Cu, Al, and O atoms, respectively.
    The Cu atoms surrounded by the O tetrahedron form a diamond lattice.
    (c) The atomic energy level diagram of the Cu$^{2+}$ ion in the tetrahedral crystal field.
  }
\label{fig:schematics}
\end{figure}

A simple atomic $t_{\rm 2g}^5$ model, in which five electrons occupying the triply degenerate $t_{\rm 2g}$-orbital are
under strong Coulomb interactions, can give a hint of how to realize the strongly correlated $J_{\rm eff}$=1/2 state,
even with small SOC. A nonzero SOC within the atomic $t_{\rm 2g}^5$ model favors the $J_{\rm eff}$=1/2 doublet as
its ground state~\cite{Abragam1970}.
Instead of considering the complicated multiplet structure composed of five electrons, the single hole in the atomic $t_{\rm 2g}^5$ model
is represented by a simple non-interacting Hamiltonian that reads $\mathcal{H}=\lambda \mathbf{l}_{\rm eff} \cdot \mathbf{s} +\Delta (l^z_{\rm eff})^2$,
where $\lambda$ is the atomic SOC and $\Delta$ is the tetragonal crystal field induced by Jahn-Teller distortion.
Note that hereafter $j_{\rm eff}$, $l_{\rm eff}$, and $s$ ($J_{\rm eff}$, $L_{\rm eff}$, and $S$) stand for single-particle (multi-particle)
total, orbital, and spin angular momenta, respectively. The lowest eigenstate of the single hole is Kramers doublet, written as
\begin{equation}
|\psi_{\pm}\rangle = \sqrt{\alpha} |l_{\rm eff}^z=0\rangle |\pm\rangle +\sqrt{1-\alpha}|l_{\rm eff}^z=\pm 1\rangle |\mp\rangle,
\end{equation}
where $|l_{\rm eff}^z=0\rangle =|d_{xy}\rangle$, $|l_{\rm eff}^z=\pm 1\rangle= -\frac{1}{\sqrt{2}}(i|d_{zx}\rangle \pm |d_{yz}\rangle)$,
and $|\pm\rangle$ denotes the spin-1/2 spinor~\cite{Jackeli2009}. Once Jahn-Teller distortion is dominant ($\Delta\gg\lambda$),
the orbital degeneracy is lifted, and the orbital angular momentum is quenched; thus, we end up with the spin-only $S$=1/2 state
($\alpha$=1) accompanied by the symmetry-lowering tetragonal distortion, which frequently occurs among 3$d$ transition metal oxides.
In the strong SOC limit or small Jahn-Teller limit, the spin-orbital-entangled $J_{\rm eff}$=1/2 state ($\alpha$=1/3) arises
while preserving the cubic symmetry. When the atomic $t_{\rm 2g}^5$ is embedded in a crystal, two limiting solutions are possible
due to the competition between the Jahn-Teller distortions and SOC [Fig.~\ref{fig:schematics}(a)].
Therefore, strong electron correlation and the narrow bandwidth of $d$-orbitals in cubic environment are a simple recipe for
the crystalline realization of the atomic $t_{\rm 2g}^5$ model, and thus, for the strongly correlated $J_{\rm eff}$=1/2 state.

In this article, we report the density-functional-theory (DFT) and dynamical mean-field theory (DMFT) calculation results to demonstrate
that the CuAl$_2$O$_4$ spinel represents the strongly correlated $J_{\rm eff}$=1/2 Mott phase by hosting the crystalline version
of the atomic $t_{\rm 2g}^5$ model. Spin-orbital entanglement in this weak SOC limit is ascribed to the tetrahedrally coordinated
$t_{\rm 2g}^5$ in the isolated CuO$_4$. Because $t_{2g}$-orbitals are not directed to the ligands in tetrahedra, the weak $d$-$p$
hybridization in CuO$_4$ reduces the energy gain from the Jahn-Teller distortions and makes the quenching of the orbital angular
momentum unlikely. And disconnected tetrahedra reduce the bandwidth of 3$d$-orbitals, approaching the atomic $t_{\rm 2g}^5$ limit.
Cooperating with large electron correlation, the $J_{\rm eff}$=1/2 ground state from the $L_{\rm eff}$=1 orbital and $S$=1/2 spin
angular momenta  are stabilized even with the small strength
of the bare SOC $\lambda_0~(\sim$50~meV) of Cu $d$-orbitals. In the strongly correlated $J_{\rm eff}$=1/2 state, many-body multiplets
and a one-particle state appear concurrently in the hole and electron excitation spectra of CuAl$_2$O$_4$, respectively.

Our total energy and electronic structure calculations were based on DFT within the PBEsol functionals~\cite{PBEsol}, as implemented in Elk code~\cite{elk}.
Brillouin zone integrations were performed using 6$\times$6$\times$6 grid sampling; the basis size was determined by $RK_{\rm max}$=9.0.
We fully optimized the structure with the force criterion of 5$\times$10$^{-4}$~eV/\AA.
The simplified rotationally invariant DFT+$U$ formalism by Dudarev {\it et al.}~\cite{dudarev} was adopted in the DFT+$U$+SOC calculations.
For the magnetic structure, we employed a collinear N\'{e}el antiferromagnetic order in which the moments were aligned along the $c$-axis.
\\

\begin{figure}
  \centering
  \includegraphics[width=0.47\textwidth,type=pdf,ext=.pdf,read=.pdf]{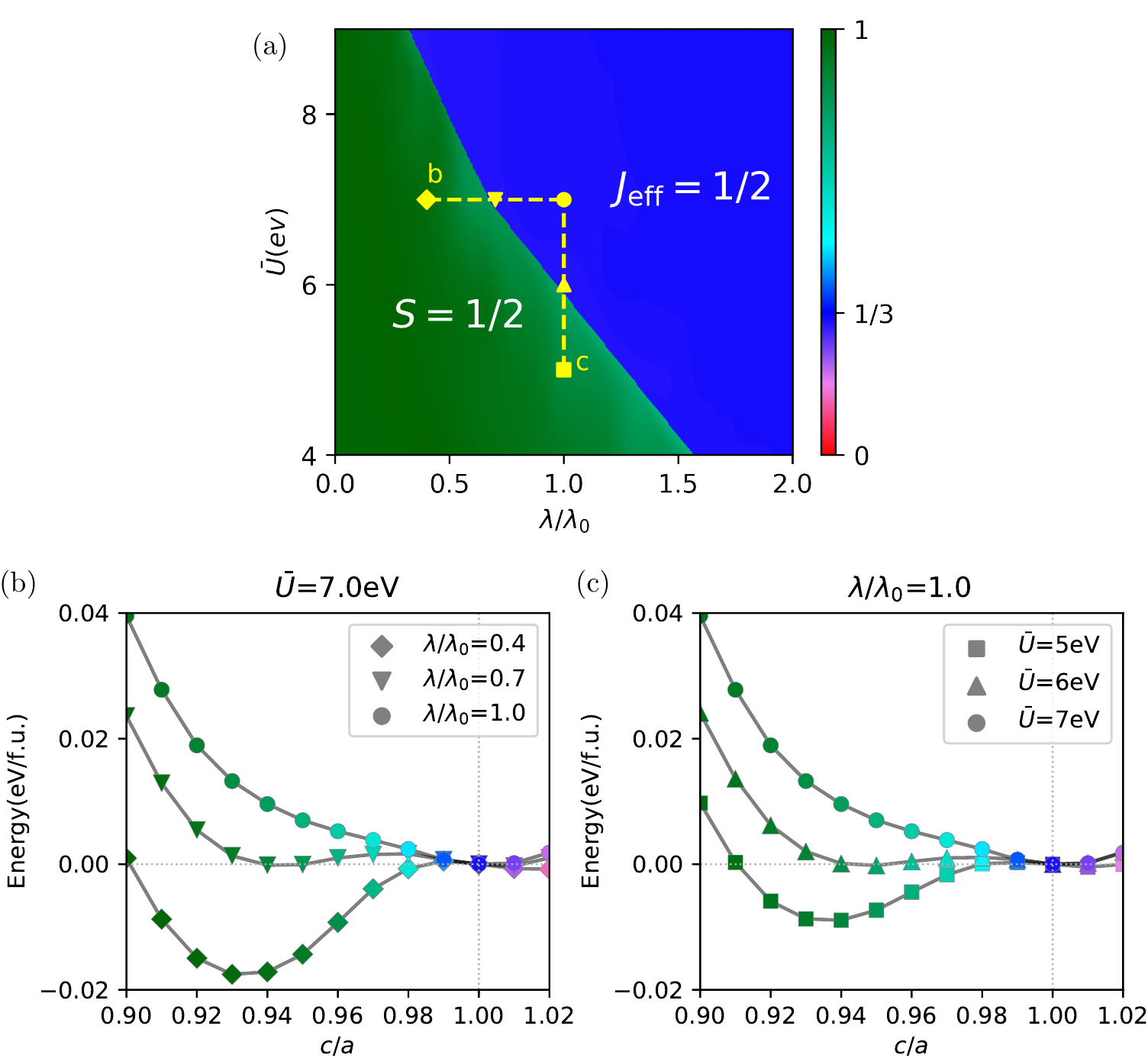}
  \caption{
Phase diagram of CuAl$_2$O$_4$ from density functional theory calculations.
(a) The phase diagram as a function of Coulomb interaction ($\bar{U}$) and spin-orbit coupling ($\lambda$).
(b),(c) Total energy curve vs $c/a$ with (b) varying $\bar{U}$, fixed $\lambda$
and (c) varying $\lambda$, fixed $\bar{U}$.
Different symbols of each energy curve indicate the corresponding parameters set in the phase diagram (a).
Colour schemes denote $\alpha$ values for given solutions.\\
}
\label{fig:phasediagram}
\end{figure}

{\it U-$\lambda$ phase diagram.}---
CuAl$_2$O$_4$ is one of the rare normal spinel cuprates with Cu$^{2+}$ at the tetrahedral site [Fig.~\ref{fig:schematics}(b)].
Recent structure analysis from x-ray and neutron powder diffraction data confirmed that it shows the cubic symmetry with $c/a=$1 (space group $Fd$-$3m$, no. 227)~\cite{Nirmala}.
In these spinel cuprates, the well-isolated CuO$_4$ tetrahedra form a diamond lattice.
In the cubic crystal field of ligand tetrahedra, the $d^9$ electrons in the Cu$^{2+}$ ion fully occupy the $e_{\rm g}$-orbitals,
leaving a single hole in the $t_{\rm 2g}$ subshell [Fig.~\ref{fig:schematics}(c)]. There is no common oxygen shared by the neighboring CuO$_4$ tetrahedra.
This drives the system closer to the atomic $t_{\rm 2g}^5$ limit, with a small $d$-orbital bandwidth and strong electron correlations. The small
energy gain from the Jahn-Teller distortion of the tetrahedron cage makes CuAl$_2$O$_4$ a promising candidate to host the $J_{\rm eff}$=1/2 state
in 3$d$ transition metal oxides.

We explored the DFT phase diagram of CuAl$_2$O$_4$ by plotting $\alpha$ defined in Eq.~(1) as a function of $\bar{U}$ and $\lambda$
[Fig.~\ref{fig:phasediagram}(a)]. For the given value of $\bar{U}$ and $\lambda$, we investigated the global minimum solution by varying
volume $V$ and tetragonal distortion $c/a$.
$\alpha$ has been extracted from the muffin tin orbital basis of a single hole wavefunction.
The phase diagram is divided into blue and green regions that correspond to
the spin-orbital-entangled $J_{\rm eff}$=1/2 ($\alpha \sim$1/3, $c/a \sim$1) and the Jahn-Teller distorted $S$=1/2 ($\alpha\sim$1, $c/a<$1) states,
respectively. The competition between SOC and Jahn-Teller distortion results in the separation of two distinct solutions. As correlation strength
increases, the phase boundary shifts toward the smaller $\lambda$, demonstrating that the SOC is enhanced effectively by electron correlation~\cite{Liu2008,Pesin2010}
and the cubic $J_{\rm eff}$=1/2 state is stabilized. In Fig.~\ref{fig:phasediagram}(b), the total energy curves are depicted with a fixed value of
$\bar{U}$ (=7~eV) and varying $\lambda$. For small SOC, two local minima appear in the total energy curves at $c/a\sim0.93$ and $c/a\sim1$,
corresponding to the $S$=1/2 and $J_{\rm eff}$=1/2 states, respectively. For nominal SOC strength ($\lambda/\lambda_0$=0.4), the $S$=1/2 state
at $c/a$=0.93 has the lowest energy. Increasing $\lambda$ stabilizes the local minimum at $c/a\sim1$ and simultaneously destabilizes the one
at $c/a<1$ , leading to a discontinuous transition of the energy minimum from tetragonal $S$=1/2 to cubic $J_{\rm eff}$=1/2 states.
Similar behavior occurs in the total energy curves with a fixed $\lambda$ (=$\lambda_0$) and varying $\bar{U}$; increasing $\bar{U}$ also tends
to make the $J_{\rm eff}$=1/2 state more stable than the $S$=1/2 state (Fig.~\ref{fig:phasediagram}c). The strong electron correlation helps
the small SOC of the Cu $d$-orbital to overcome the Jahn-Teller distortion, enabling the spin-orbital-entangled ground state.

A reasonable value of the correlation strength could be estimated by Cococcioni's linear response approach~\cite{Cococcioni2005}.
In this approach, the response function is $\chi=\frac{\partial n}{\partial \mu}$ where $\mu$ is the potential shift and $n$ is the number of
electrons on Hubbard atom. The effective interaction parameter $\bar{U}$ can be obtained by inverting the self-consistent response function and
subtracting out the bare (non-interacting) response:
\begin{equation}
\label{linearresponse}
\bar{U}= (\chi^{-1}_0 - \chi^{-1})
\end{equation}
We obtained $\bar{U}\sim$ 9~eV for Cu 3$d$-orbitals within this formalism. From the phase diagram,
critical value of $\bar{U}$ for the $J_{\rm eff}$=1/2 state is about 6~eV,
thereby, the $J_{\rm eff}$=1/2 state could be a plausible ground state of CuAl$_2$O$_4$.

\onecolumngrid

\begin{figure}[h]
  \centering
  \includegraphics[width=0.95\textwidth,type=pdf,ext=.pdf,read=.pdf]{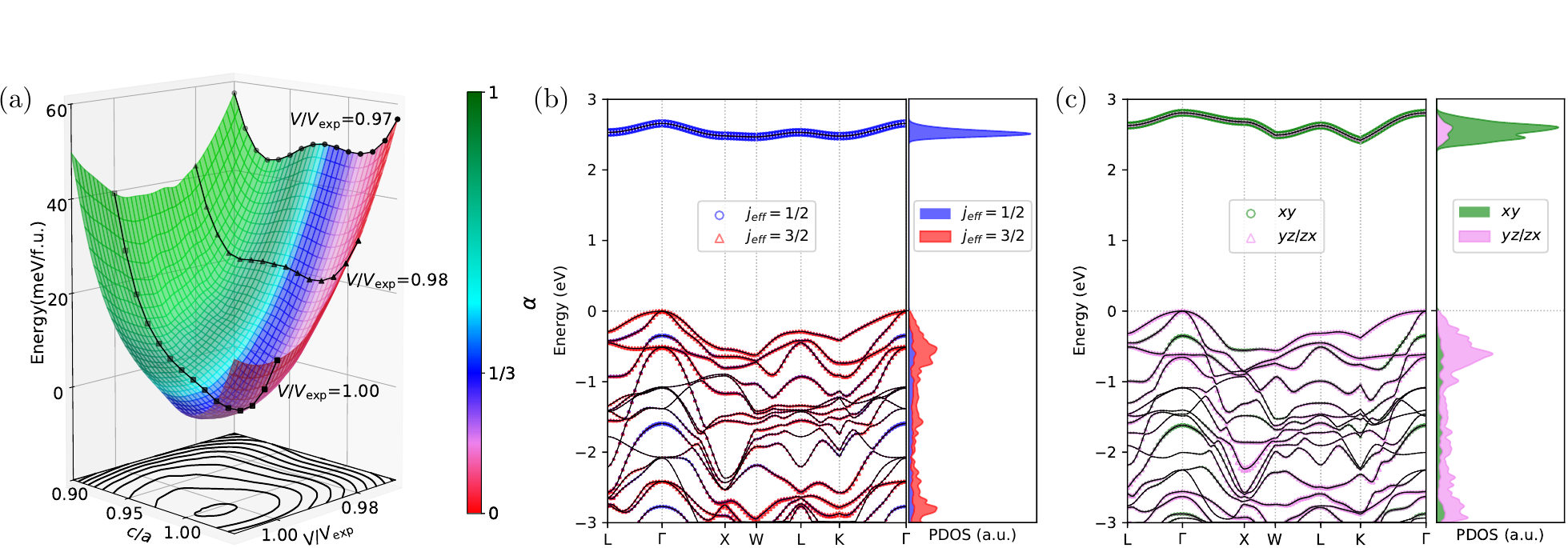}
  \caption{
DFT total energy landscape and two competing phases.
(a) Total energy landscape as a function of $V/V_{\rm exp}$ and $c/a$ with $U$=7eV and $\lambda/\lambda_0$=1.
(b),(c) Band structure and projected density-of-states (PDOS) for (b) $V/V_{\rm exp}$=1.0025 and $c/a$=1.00 
and (c) $V/V_{\rm exp}$=0.970 and $c/a$=0.93, corresponding to $J_{\rm eff}$=1/2 and $S$=1/2 states, respectively.
}
\label{fig:energyband}
\end{figure}

\twocolumngrid

{\it Total energy landscape.}---
For $\bar{U}$=7eV and $\lambda/\lambda_0$=1, we have investigated the total energy landscape as a function of $V/V_{\rm exp}$ and $c/a$.
As shown in Fig.~\ref{fig:energyband}(a), the only stable (and thus global) minimum solution occurs at
$V/V_{\rm exp}$=1.0025 and $c/a$=1, whose structural properties are consistent with the previous experimental results~\cite{Nirmala}.
The electronic structure and projected density-of-state (PDOS) at $V/V_{\rm exp}$=1, $c/a$=1 is shown in Fig.~\ref{fig:energyband}(b).
In the band structure, the unoccupied band above the Fermi level can be perfectly projected onto the $j_{\rm eff}$=1/2 doublet with $\alpha$=0.32.
Since the unoccupied state in the $t_{\rm 2g}^5$ configuration basically represents a single hole, the electron-addition spectra are
well described by the spin-orbital-entangled doublet. On the other hand, the electron-removal spectra form a many-body multiplet structure,
resulting in the mixture of $j_{\rm eff}$=1/2 and 3/2 components in the PDOS plot. This differs from the common expectation for the weakly correlated
$J_{\rm eff}$=1/2 state, for example, realized in Sr$_2$IrO$_4$. The multiplet effects appearing in the electron spectrum of CuAl$_2$O$_4$ become
clear in the DMFT calculations shown later.

Even though there is no other stable solution, the total energy landscape interestingly suggests that a possible Jahn-Teller distorted $S$=1/2
state might be stabilized under high pressure. At higher pressure, the Cu-O bond length gets shorter, giving rise to larger crystal field
splittings induced by Jahn-Teller distortions. By constraining the volume decreased by 3\%, the $S$=1/2 state at $c/a$=0.93 has a lower energy
than the $J_{\rm eff}$=1/2 state at $c/a$=1. Therefore, the two distinct $J_{\rm eff}$=1/2 and $S$=1/2 phases can be realized with the same
sample by applying pressure values of experimentally accessible range. The electronic structure of the Jahn-Teller distorted $S$=1/2 state at
$V/V_{\rm exp}$=0.97, $c/a$=0.93 is shown in Fig.~\ref{fig:energyband} (c). Due to the large tetragonal distortion, the single hole spectrum
of the unoccupied $t_{\rm 2g}$ bands is mostly composed of $d_{xy}$-orbital with $\alpha$=0.89.
\\

\begin{figure}
  \centering
  \includegraphics[width=0.50\textwidth]{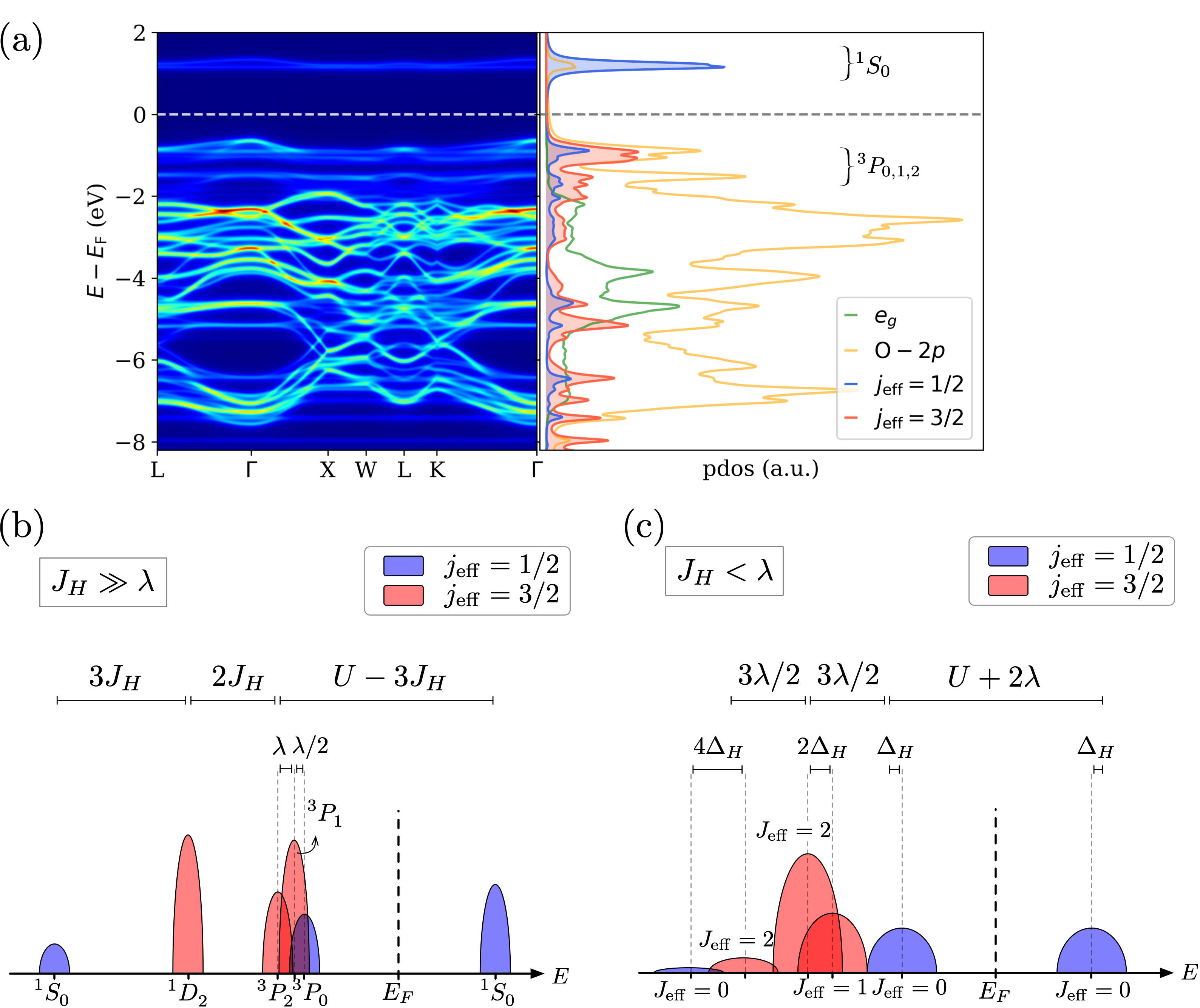}
  \caption{
Multiplets in dynamical mean field theory (DMFT) calculations.
(a) Spectral weights and PDOS from DMFT calculations for $U$=8~eV, $J_H$=1~eV, $\lambda$=0.05~eV.
While the spectral gap is roughly proportional to $U$, the splitting of the hole spectra below the Fermi level depends on $\lambda$ and $J_{\rm H}$.
Schematic illustration of the single-electron/hole excitation spectra from (b), the strongly correlated ($J_{\rm H} \gg \lambda$)
and (c), the weakly correlated ($J_{\rm H} < \lambda$) $J_{\rm eff}$=1/2 ground state.
In (c), $\Delta_{\rm H} = 3J_{\rm H}/2$ is used for simplicity.
}
\label{fig:dmft}
\end{figure}

{\it DMFT calculations.}---
We also conducted DMFT calculations on top of the DFT-based Wannier Hamiltonian to clarify how robust the $J_{\rm eff}$-ness is under quantum fluctuations.
Maximally localized Wannier functions~\cite{MLWF} were obtained from the DFT full Cu 3$d$+oxygen 2$p$ bands in the absence of $U$ and SOC.
As such, SOC and the rotationally invariant local Coulomb interaction at each Cu ion were treated by DMFT,
where the double counting correction was applied using the fully localized limit scheme~\cite{Solovyev1994}.
The correlations involving $e_{\rm g}$-orbitals were calculated by the Hartree-Fock approximation and the oxygen orbitals were assumed to be noninteracting~\cite{Park2012,Haule2014}.
We employed the exact diagonalization (ED)~\cite{dmft_ed} as an impurity solver for the zero-temperature DMFT calculations.

We present the DMFT spectral function and PDOS in Fig.~\ref{fig:dmft}(a) for a realistic parameter set ($U$=8~eV,
$J_{\rm H}$=1~eV, and $\lambda$=50~meV). First of all, we note that the strong $j_{\rm eff}$=1/2 hole character is also manifested in the DMFT
calculation, indicating that the $J_{\rm eff}$=1/2 state is stable with respect to local quantum fluctuations. The states below the Fermi level
exhibit an additional dynamic weight transfer originating from multiplets focusing on the $t_{\rm 2g}$ manifold just below the Fermi level;
the lowest $t_{\rm 2g}^5 \rightarrow t_{\rm 2g}^4$ excitation spectra show a mixture of $j_{\rm eff}$=1/2 and 3/2 characters. This reveals
a unique signature of the strongly correlated $J_{\rm eff}$=1/2 state obeying the $LS$-coupling scheme, which is distinct from the weakly correlated
counterpart such as Sr$_2$IrO$_4$ close to the $jj$-coupling regime.

The weight distribution can be understood by the  atomic $t_{\rm 2g}^5$ model with dominating Hund's coupling $J_{\rm H} \gg \lambda$
in Fig.~\ref{fig:dmft}(b). In the atomic model, the lowest peak below the Fermi level is composed of three overlapping sub-peak structures, denoted by
$^3P_0, ^3P_1$, and $^3P_2$. Each sub-peak is categorized by either $j_{\rm eff}$=1/2 ($^3P_0$) or $j_{\rm eff}$=3/2 ($^3P_1, ^3P_2$); the mixture
of the $j_{\rm eff}$=1/2 and 3/2 components in the lowest hole excitation shows the close correspondence between the DMFT spectral function and
the atomic multiplet description. (This behaviour becomes even clearer in an independent $t_{\rm 2g}$-only DMFT calculation, excluding the $e_{\rm g}$
and oxygen contribution as shown in Supplementary Information.) The uniqueness of the excitation spectra is further highlighted by comparison with
the case of iridates. We investigated the $t_{\rm 2g}^5$ atomic model with a strong SOC regime ($\lambda > J_{\rm H}$) in Fig.~\ref{fig:dmft}(c)
that can be compared to the $J_{\rm eff}$=1/2 state in 5$d$ iridates~\cite{Kim2008}. In this strong SOC regime closer to the $jj$-coupling scheme,
electron removal spectra exhibit two prominent peaks, clearly separated by the large SOC and categorized by $j_{\rm eff}$=1/2 and $j_{\rm eff}$=3/2 character,
respectively. This feature is reflected in the previous experimental and theoretical reports in Sr$_2$IrO$_4$~\cite{Kim2008,Martins2016,Arita2012,Zhang2013,Parschke2017},
where the DFT single-particle band structure provides a reasonable description given that multiplet effects are less important in this
parameter range. (See Supplementary Information for the hole excitation spectrum of the atomic $t_{\rm 2g}^5$ model over the whole parameter range.)

Under the cubic symmetry, the SOC puts a single hole in the $t_{\rm 2g}^5$ configuration into the $j_{\rm eff}$=1/2 state
to lower the energy and therefore the ground state becomes $J_{\rm eff}$=1/2.
As a result, the $J_{\rm eff}$=1/2 ground state is represented by the unoccupied $j_{\rm eff}$=1/2 state in the band structure.
But the occupied spectrum of the $J_{\rm eff}$=1/2 states in $jj$- and $LS$-coupling regime behave very differently from each other.
In the $jj$-coupling scheme ($\lambda> J_H$), the occupied states are well described by the single particle picture.
Then we can see the clear separation between $j_{\rm eff}$=1/2 and $j_{\rm eff}$=3/2 states of the occupied bands as previously shown in iridates.
On the other hand, however, the single particle description is no longer valid in the $LS$-coupling scheme ($\lambda\ll J_H$) to explain the occupied spectrum,
and thus $t_{\rm 2g}^4$ multiplet structures are inevitable.
This is the uniqueness of the newly emerging ``strongly correlated'' $J_{\rm eff}$=1/2 state of CuAl$_2$O$_4$ in which the occupied spectrum is governed by the $LS$-coupling scheme.

Recently, Nirmala and coworkers reported the magnetic susceptibility as well as heat capacity data and found no signature of long range magnetic order down to 0.4~K.
The DMFT calculations show a genuine Mott insulator without breaking the time-reversal symmetry, whereas the DFT solution requires symmetry breaking to open a gap in the primitive unit cell calculations.
Although the copper network in CuAl$_2$O$_4$ has a bipartite structure, the paramagnetic ground state persists in the DMFT results.
The hole weights are equally distributed in the Kramers pair in Eq.~(1) for the entire parameter range considered in this DMFT study.
Even if we apply a small staggered magnetic field to stabilize an antiferromagnetic order, the magnetic moment quickly disappears as soon as the staggered field is turned off.
The suppression of magnetic order may arise from frustration effects, stemming from larger second-neighbor hopping
amplitudes than nearest-neighbor ones~\cite{Bergman2007}.(See Supplementary Information.)
The origin and nature of the nonmagnetic Mott phase of CuAl$_2$O$_4$ are beyond the scope of the present work.
Given the possibility of being extended to the $J_{\rm eff}$=1/2 spin glass or liquid phase, however, the lack of long-range magnetic order is of great interest, requiring further study.

{\it Remarks.}---
A sizable amount of the site disorder between Cu and Al has been recently reported in powder samples of CuAl$_2$O$_4$~\cite{Nirmala}. 
To check the robustness of $J_{\rm eff}$=1/2 picture under disorder, we performed the DFT calculation with containing 50\% site disorder (see Supplementary Information).
Even under the maximal disorder, the single hole at the tetrahedral Cu site preserves the $j_{\rm eff}$=1/2 character. As shown in Supplementary Information Fig.~5S,
two separated bands appear above the Fermi level,
which correspond to the unoccupied Cu $d$-orbitals from each tetrahedral and octahedral site.
The lower band is perfectly projected onto the $j_{\rm eff}$=1/2 state in the tetrahedral site, whereas the higher one comes from the $e_{\rm g}$ state in the octahedral site.
It indicates that the localized unoccupied states in each
tetrahedral and octahedral site behave almost independently, indeed manifesting the $J_{\rm eff}$-ness of the tetrahedral Cu$^{2+}$ even with the significant amount of disorder.
To understand spin glass behaviour shown in the powder sample,
magnetic interactions under the mixture of tetrahedral site $J_{\rm eff}$=1/2 and octahedral site $e_g$ states should be studied.

{\it Conclusion.}---
We have shown the theoretical evidences that CuAl$_2$O$_4$ spinel is a strongly correlated $J_{\rm eff}$=1/2 Mott insulator.
The first-principles total energy calculations reproduce the previous X-ray data reporting cubic structure of CuAl$_2$O$_4$.
And its band structure clearly shows that the unoccupied band is well characterized by the $j_{\rm eff}$=1/2 state.
The DMFT calculations uncover the uniqueness of excitation spectra of a strongly correlated $J_{\rm eff}$=1/2 Mott phase in CuAl$_2$O$_4$,
realizing a $J_{\rm eff}$=1/2 state in the $LS$-coupling limit.
\\

This work was supported by Institute for Basic Science (IBS) in Korea (Grant No. IBS-R009-D1 (CHK, SB), IBS-R024-D1 (AG), IBS-R009-G1 (HC, JGP)), the Basic Science Research Program of the National Research Foundation (NRF) of Korea under Grant No. 2016R1D1A1B03933255, 2017M3D1A1040828 and 2019R1A2C1010498 (HJ).
The work of GVV and SVS was supported by the Russian science foundation (grant 17-12-01207), while DIKh thanks the Deutsche Forschungsgemeinschaft (SFB 1238) and German Excellence Initiative.
\\


\begin{thebibliography}{39}%
\makeatletter
\providecommand \@ifxundefined [1]{%
 \@ifx{#1\undefined}
}%
\providecommand \@ifnum [1]{%
 \ifnum #1\expandafter \@firstoftwo
 \else \expandafter \@secondoftwo
 \fi
}%
\providecommand \@ifx [1]{%
 \ifx #1\expandafter \@firstoftwo
 \else \expandafter \@secondoftwo
 \fi
}%
\providecommand \natexlab [1]{#1}%
\providecommand \enquote  [1]{``#1''}%
\providecommand \bibnamefont  [1]{#1}%
\providecommand \bibfnamefont [1]{#1}%
\providecommand \citenamefont [1]{#1}%
\providecommand \href@noop [0]{\@secondoftwo}%
\providecommand \href [0]{\begingroup \@sanitize@url \@href}%
\providecommand \@href[1]{\@@startlink{#1}\@@href}%
\providecommand \@@href[1]{\endgroup#1\@@endlink}%
\providecommand \@sanitize@url [0]{\catcode `\\12\catcode `\$12\catcode
  `\&12\catcode `\#12\catcode `\^12\catcode `\_12\catcode `\%12\relax}%
\providecommand \@@startlink[1]{}%
\providecommand \@@endlink[0]{}%
\providecommand \url  [0]{\begingroup\@sanitize@url \@url }%
\providecommand \@url [1]{\endgroup\@href {#1}{\urlprefix }}%
\providecommand \urlprefix  [0]{URL }%
\providecommand \Eprint [0]{\href }%
\providecommand \doibase [0]{http://dx.doi.org/}%
\providecommand \selectlanguage [0]{\@gobble}%
\providecommand \bibinfo  [0]{\@secondoftwo}%
\providecommand \bibfield  [0]{\@secondoftwo}%
\providecommand \translation [1]{[#1]}%
\providecommand \BibitemOpen [0]{}%
\providecommand \bibitemStop [0]{}%
\providecommand \bibitemNoStop [0]{.\EOS\space}%
\providecommand \EOS [0]{\spacefactor3000\relax}%
\providecommand \BibitemShut  [1]{\csname bibitem#1\endcsname}%
\let\auto@bib@innerbib\@empty
\bibitem [{\citenamefont {Imada}\ \emph {et~al.}(1998)\citenamefont {Imada},
  \citenamefont {Fujimori},\ and\ \citenamefont {Tokura}}]{Imada1998}%
  \BibitemOpen
  \bibfield  {author} {\bibinfo {author} {\bibfnamefont {M.}~\bibnamefont
  {Imada}}, \bibinfo {author} {\bibfnamefont {A.}~\bibnamefont {Fujimori}}, \
  and\ \bibinfo {author} {\bibfnamefont {Y.}~\bibnamefont {Tokura}},\ }\href
  {\doibase 10.1103/RevModPhys.70.1039} {\bibfield  {journal} {\bibinfo
  {journal} {Rev. Mod. Phys.}\ }\textbf {\bibinfo {volume} {70}},\ \bibinfo
  {pages} {1039} (\bibinfo {year} {1998})}\BibitemShut {NoStop}%
\bibitem [{\citenamefont {Kugel}\ and\ \citenamefont
  {Khomski}(1982)}]{Kugel1982}%
  \BibitemOpen
  \bibfield  {author} {\bibinfo {author} {\bibfnamefont {K.~I.}\ \bibnamefont
  {Kugel}}\ and\ \bibinfo {author} {\bibfnamefont {D.~I.}\ \bibnamefont
  {Khomski}},\ }\href {http://stacks.iop.org/0038-5670/25/i=4/a=R03} {\bibfield
   {journal} {\bibinfo  {journal} {Sov. Phys. Usp.}\ }\textbf {\bibinfo
  {volume} {25}},\ \bibinfo {pages} {231} (\bibinfo {year} {1982})}\BibitemShut
  {NoStop}%
\bibitem [{\citenamefont {Khomskii}(2014)}]{Khomskii2014}%
  \BibitemOpen
  \bibfield  {author} {\bibinfo {author} {\bibfnamefont {D.~I.}\ \bibnamefont
  {Khomskii}},\ }\href@noop {} {\emph {\bibinfo {title} {Transition Metal
  Compounds}}}\ (\bibinfo  {publisher} {Cambridge University Press},\ \bibinfo
  {year} {2014})\BibitemShut {NoStop}%
\bibitem [{\citenamefont {Kim}\ \emph {et~al.}(2008)\citenamefont {Kim},
  \citenamefont {Jin}, \citenamefont {Moon}, \citenamefont {Kim}, \citenamefont
  {Park}, \citenamefont {Leem}, \citenamefont {Yu}, \citenamefont {Noh},
  \citenamefont {Kim}, \citenamefont {Oh}, \citenamefont {Park}, \citenamefont
  {Durairaj}, \citenamefont {Cao},\ and\ \citenamefont {Rotenberg}}]{Kim2008}%
  \BibitemOpen
  \bibfield  {author} {\bibinfo {author} {\bibfnamefont {B.~J.}\ \bibnamefont
  {Kim}}, \bibinfo {author} {\bibfnamefont {H.}~\bibnamefont {Jin}}, \bibinfo
  {author} {\bibfnamefont {S.}~\bibnamefont {Moon}}, \bibinfo {author}
  {\bibfnamefont {J.-Y.}\ \bibnamefont {Kim}}, \bibinfo {author} {\bibfnamefont
  {B.-G.}\ \bibnamefont {Park}}, \bibinfo {author} {\bibfnamefont
  {C.}~\bibnamefont {Leem}}, \bibinfo {author} {\bibfnamefont {J.}~\bibnamefont
  {Yu}}, \bibinfo {author} {\bibfnamefont {T.}~\bibnamefont {Noh}}, \bibinfo
  {author} {\bibfnamefont {C.}~\bibnamefont {Kim}}, \bibinfo {author}
  {\bibfnamefont {S.-J.}\ \bibnamefont {Oh}}, \bibinfo {author} {\bibfnamefont
  {J.-H.}\ \bibnamefont {Park}}, \bibinfo {author} {\bibfnamefont
  {V.}~\bibnamefont {Durairaj}}, \bibinfo {author} {\bibfnamefont
  {G.}~\bibnamefont {Cao}}, \ and\ \bibinfo {author} {\bibfnamefont
  {E.}~\bibnamefont {Rotenberg}},\ }\href {\doibase
  10.1103/PhysRevLett.101.076402} {\bibfield  {journal} {\bibinfo  {journal}
  {Phys. Rev. Lett.}\ }\textbf {\bibinfo {volume} {101}},\ \bibinfo {pages}
  {076402} (\bibinfo {year} {2008})}\BibitemShut {NoStop}%
\bibitem [{\citenamefont {Kim}\ \emph {et~al.}(2009)\citenamefont {Kim},
  \citenamefont {Ohsumi}, \citenamefont {Komesu}, \citenamefont {Sakai},
  \citenamefont {Morita}, \citenamefont {Takagi},\ and\ \citenamefont
  {Arima}}]{Kim2009c}%
  \BibitemOpen
  \bibfield  {author} {\bibinfo {author} {\bibfnamefont {B.~J.}\ \bibnamefont
  {Kim}}, \bibinfo {author} {\bibfnamefont {H.}~\bibnamefont {Ohsumi}},
  \bibinfo {author} {\bibfnamefont {T.}~\bibnamefont {Komesu}}, \bibinfo
  {author} {\bibfnamefont {S.}~\bibnamefont {Sakai}}, \bibinfo {author}
  {\bibfnamefont {T.}~\bibnamefont {Morita}}, \bibinfo {author} {\bibfnamefont
  {H.}~\bibnamefont {Takagi}}, \ and\ \bibinfo {author} {\bibfnamefont
  {T.}~\bibnamefont {Arima}},\ }\href
  {http://www.sciencemag.org/content/323/5919/1329.short} {\bibfield  {journal}
  {\bibinfo  {journal} {Science}\ }\textbf {\bibinfo {volume} {323}},\ \bibinfo
  {pages} {1329} (\bibinfo {year} {2009})}\BibitemShut {NoStop}%
\bibitem [{\citenamefont {Plumb}\ \emph {et~al.}(2014)\citenamefont {Plumb},
  \citenamefont {Clancy}, \citenamefont {Sandilands}, \citenamefont {Shankar},
  \citenamefont {Hu}, \citenamefont {Burch}, \citenamefont {Kee},\ and\
  \citenamefont {Kim}}]{Plumb2014}%
  \BibitemOpen
  \bibfield  {author} {\bibinfo {author} {\bibfnamefont {K.~W.}\ \bibnamefont
  {Plumb}}, \bibinfo {author} {\bibfnamefont {J.~P.}\ \bibnamefont {Clancy}},
  \bibinfo {author} {\bibfnamefont {L.~J.}\ \bibnamefont {Sandilands}},
  \bibinfo {author} {\bibfnamefont {V.~V.}\ \bibnamefont {Shankar}}, \bibinfo
  {author} {\bibfnamefont {Y.~F.}\ \bibnamefont {Hu}}, \bibinfo {author}
  {\bibfnamefont {K.~S.}\ \bibnamefont {Burch}}, \bibinfo {author}
  {\bibfnamefont {H.-Y.}\ \bibnamefont {Kee}}, \ and\ \bibinfo {author}
  {\bibfnamefont {Y.-J.}\ \bibnamefont {Kim}},\ }\href {\doibase
  10.1103/PhysRevB.90.041112} {\bibfield  {journal} {\bibinfo  {journal} {Phys.
  Rev. B}\ }\textbf {\bibinfo {volume} {90}},\ \bibinfo {pages} {041112}
  (\bibinfo {year} {2014})}\BibitemShut {NoStop}%
\bibitem [{\citenamefont {Kim}\ \emph {et~al.}(2014{\natexlab{a}})\citenamefont
  {Kim}, \citenamefont {Im}, \citenamefont {Han},\ and\ \citenamefont
  {Jin}}]{Kim2014}%
  \BibitemOpen
  \bibfield  {author} {\bibinfo {author} {\bibfnamefont {H.-S.}\ \bibnamefont
  {Kim}}, \bibinfo {author} {\bibfnamefont {J.}~\bibnamefont {Im}}, \bibinfo
  {author} {\bibfnamefont {M.~J.}\ \bibnamefont {Han}}, \ and\ \bibinfo
  {author} {\bibfnamefont {H.}~\bibnamefont {Jin}},\ }\href {\doibase
  10.1038/ncomms4988} {\bibfield  {journal} {\bibinfo  {journal} {Nature
  communications}\ }\textbf {\bibinfo {volume} {5}},\ \bibinfo {pages} {3988}
  (\bibinfo {year} {2014}{\natexlab{a}})}\BibitemShut {NoStop}%
\bibitem [{\citenamefont {Jackeli}\ and\ \citenamefont
  {Khaliullin}(2009)}]{Jackeli2009}%
  \BibitemOpen
  \bibfield  {author} {\bibinfo {author} {\bibfnamefont {G.}~\bibnamefont
  {Jackeli}}\ and\ \bibinfo {author} {\bibfnamefont {G.}~\bibnamefont
  {Khaliullin}},\ }\href {\doibase 10.1103/PhysRevLett.102.017205} {\bibfield
  {journal} {\bibinfo  {journal} {Phys. Rev. Lett.}\ }\textbf {\bibinfo
  {volume} {102}},\ \bibinfo {pages} {017205} (\bibinfo {year}
  {2009})}\BibitemShut {NoStop}%
\bibitem [{\citenamefont {Wang}\ and\ \citenamefont
  {Senthil}(2011)}]{Wang2011d}%
  \BibitemOpen
  \bibfield  {author} {\bibinfo {author} {\bibfnamefont {F.}~\bibnamefont
  {Wang}}\ and\ \bibinfo {author} {\bibfnamefont {T.}~\bibnamefont {Senthil}},\
  }\href {\doibase 10.1103/PhysRevLett.106.136402} {\bibfield  {journal}
  {\bibinfo  {journal} {Phys. Rev. Lett.}\ }\textbf {\bibinfo {volume} {106}},\
  \bibinfo {pages} {136402} (\bibinfo {year} {2011})}\BibitemShut {NoStop}%
\bibitem [{\citenamefont {Kim}\ \emph {et~al.}(2014{\natexlab{b}})\citenamefont
  {Kim}, \citenamefont {Krupin}, \citenamefont {Denlinger}, \citenamefont
  {Bostwick}, \citenamefont {Rotenberg}, \citenamefont {Zhao}, \citenamefont
  {Mitchell}, \citenamefont {Allen},\ and\ \citenamefont {Kim}}]{Kim2014d}%
  \BibitemOpen
  \bibfield  {author} {\bibinfo {author} {\bibfnamefont {Y.~K.}\ \bibnamefont
  {Kim}}, \bibinfo {author} {\bibfnamefont {O.}~\bibnamefont {Krupin}},
  \bibinfo {author} {\bibfnamefont {J.~D.}\ \bibnamefont {Denlinger}}, \bibinfo
  {author} {\bibfnamefont {A.}~\bibnamefont {Bostwick}}, \bibinfo {author}
  {\bibfnamefont {E.}~\bibnamefont {Rotenberg}}, \bibinfo {author}
  {\bibfnamefont {Q.}~\bibnamefont {Zhao}}, \bibinfo {author} {\bibfnamefont
  {J.~F.}\ \bibnamefont {Mitchell}}, \bibinfo {author} {\bibfnamefont {J.~W.}\
  \bibnamefont {Allen}}, \ and\ \bibinfo {author} {\bibfnamefont {B.~J.}\
  \bibnamefont {Kim}},\ }\href {\doibase 10.1126/science.1251151} {\bibfield
  {journal} {\bibinfo  {journal} {Science}\ }\textbf {\bibinfo {volume}
  {345}},\ \bibinfo {pages} {187} (\bibinfo {year}
  {2014}{\natexlab{b}})}\BibitemShut {NoStop}%
\bibitem [{\citenamefont {Shitade}\ \emph {et~al.}(2009)\citenamefont
  {Shitade}, \citenamefont {Katsura}, \citenamefont {Kune{\v{s}}},
  \citenamefont {Qi}, \citenamefont {Zhang},\ and\ \citenamefont
  {Nagaosa}}]{Shitade2009}%
  \BibitemOpen
  \bibfield  {author} {\bibinfo {author} {\bibfnamefont {A.}~\bibnamefont
  {Shitade}}, \bibinfo {author} {\bibfnamefont {H.}~\bibnamefont {Katsura}},
  \bibinfo {author} {\bibfnamefont {J.}~\bibnamefont {Kune{\v{s}}}}, \bibinfo
  {author} {\bibfnamefont {X.-L.}\ \bibnamefont {Qi}}, \bibinfo {author}
  {\bibfnamefont {S.-C.}\ \bibnamefont {Zhang}}, \ and\ \bibinfo {author}
  {\bibfnamefont {N.}~\bibnamefont {Nagaosa}},\ }\href {\doibase
  10.1103/PhysRevLett.102.256403} {\bibfield  {journal} {\bibinfo  {journal}
  {Phys. Rev. Lett.}\ }\textbf {\bibinfo {volume} {102}},\ \bibinfo {pages}
  {256403} (\bibinfo {year} {2009})}\BibitemShut {NoStop}%
\bibitem [{\citenamefont {Kim}\ \emph {et~al.}(2012)\citenamefont {Kim},
  \citenamefont {Kim}, \citenamefont {Jeong}, \citenamefont {Jin},\ and\
  \citenamefont {Yu}}]{Kim2012}%
  \BibitemOpen
  \bibfield  {author} {\bibinfo {author} {\bibfnamefont {C.~H.}\ \bibnamefont
  {Kim}}, \bibinfo {author} {\bibfnamefont {H.~S.}\ \bibnamefont {Kim}},
  \bibinfo {author} {\bibfnamefont {H.}~\bibnamefont {Jeong}}, \bibinfo
  {author} {\bibfnamefont {H.}~\bibnamefont {Jin}}, \ and\ \bibinfo {author}
  {\bibfnamefont {J.}~\bibnamefont {Yu}},\ }\href {\doibase
  10.1103/PhysRevLett.108.106401} {\bibfield  {journal} {\bibinfo  {journal}
  {Phys. Rev. Lett.}\ }\textbf {\bibinfo {volume} {108}},\ \bibinfo {pages}
  {106401} (\bibinfo {year} {2012})}\BibitemShut {NoStop}%
\bibitem [{\citenamefont {Chaloupka}\ \emph {et~al.}(2010)\citenamefont
  {Chaloupka}, \citenamefont {Jackeli},\ and\ \citenamefont
  {Khaliullin}}]{Chaloupka2010}%
  \BibitemOpen
  \bibfield  {author} {\bibinfo {author} {\bibfnamefont {J.}~\bibnamefont
  {Chaloupka}}, \bibinfo {author} {\bibfnamefont {G.}~\bibnamefont {Jackeli}},
  \ and\ \bibinfo {author} {\bibfnamefont {G.}~\bibnamefont {Khaliullin}},\
  }\href {\doibase 10.1103/PhysRevLett.105.027204} {\bibfield  {journal}
  {\bibinfo  {journal} {Phys. Rev. Lett.}\ }\textbf {\bibinfo {volume} {105}},\
  \bibinfo {pages} {027204} (\bibinfo {year} {2010})}\BibitemShut {NoStop}%
\bibitem [{\citenamefont {Kitagawa}\ \emph {et~al.}(2018)\citenamefont
  {Kitagawa}, \citenamefont {Takayama}, \citenamefont {Matsumoto},
  \citenamefont {Kato}, \citenamefont {Takano}, \citenamefont {Kishimoto},
  \citenamefont {Bette}, \citenamefont {Dinnebier}, \citenamefont {Jackeli},\
  and\ \citenamefont {Takagi}}]{Kitagawa2018}%
  \BibitemOpen
  \bibfield  {author} {\bibinfo {author} {\bibfnamefont {K.}~\bibnamefont
  {Kitagawa}}, \bibinfo {author} {\bibfnamefont {T.}~\bibnamefont {Takayama}},
  \bibinfo {author} {\bibfnamefont {Y.}~\bibnamefont {Matsumoto}}, \bibinfo
  {author} {\bibfnamefont {A.}~\bibnamefont {Kato}}, \bibinfo {author}
  {\bibfnamefont {R.}~\bibnamefont {Takano}}, \bibinfo {author} {\bibfnamefont
  {Y.}~\bibnamefont {Kishimoto}}, \bibinfo {author} {\bibfnamefont
  {S.}~\bibnamefont {Bette}}, \bibinfo {author} {\bibfnamefont
  {R.}~\bibnamefont {Dinnebier}}, \bibinfo {author} {\bibfnamefont
  {G.}~\bibnamefont {Jackeli}}, \ and\ \bibinfo {author} {\bibfnamefont
  {H.}~\bibnamefont {Takagi}},\ }\href {\doibase 10.1038/nature25482}
  {\bibfield  {journal} {\bibinfo  {journal} {Nature}\ }\textbf {\bibinfo
  {volume} {554}},\ \bibinfo {pages} {341} (\bibinfo {year}
  {2018})}\BibitemShut {NoStop}%
\bibitem [{\citenamefont {Winter}\ \emph {et~al.}(2017)\citenamefont {Winter},
  \citenamefont {Tsirlin}, \citenamefont {Daghofer}, \citenamefont {van~den
  Brink}, \citenamefont {Singh}, \citenamefont {Gegenwart},\ and\ \citenamefont
  {Valent{\'{\i}}}}]{Winter2017}%
  \BibitemOpen
  \bibfield  {author} {\bibinfo {author} {\bibfnamefont {S.~M.}\ \bibnamefont
  {Winter}}, \bibinfo {author} {\bibfnamefont {A.~A.}\ \bibnamefont {Tsirlin}},
  \bibinfo {author} {\bibfnamefont {M.}~\bibnamefont {Daghofer}}, \bibinfo
  {author} {\bibfnamefont {J.}~\bibnamefont {van~den Brink}}, \bibinfo {author}
  {\bibfnamefont {Y.}~\bibnamefont {Singh}}, \bibinfo {author} {\bibfnamefont
  {P.}~\bibnamefont {Gegenwart}}, \ and\ \bibinfo {author} {\bibfnamefont
  {R.}~\bibnamefont {Valent{\'{\i}}}},\ }\href {\doibase
  10.1088/1361-648x/aa8cf5} {\bibfield  {journal} {\bibinfo  {journal} {Journal
  of Physics: Condensed Matter}\ }\textbf {\bibinfo {volume} {29}},\ \bibinfo
  {pages} {493002} (\bibinfo {year} {2017})}\BibitemShut {NoStop}%
\bibitem [{\citenamefont {Wan}\ \emph {et~al.}(2011)\citenamefont {Wan},
  \citenamefont {Turner}, \citenamefont {Vishwanath},\ and\ \citenamefont
  {Savrasov}}]{Wan2011}%
  \BibitemOpen
  \bibfield  {author} {\bibinfo {author} {\bibfnamefont {X.}~\bibnamefont
  {Wan}}, \bibinfo {author} {\bibfnamefont {A.~M.}\ \bibnamefont {Turner}},
  \bibinfo {author} {\bibfnamefont {A.}~\bibnamefont {Vishwanath}}, \ and\
  \bibinfo {author} {\bibfnamefont {S.~Y.}\ \bibnamefont {Savrasov}},\ }\href
  {\doibase 10.1103/PhysRevB.83.205101} {\bibfield  {journal} {\bibinfo
  {journal} {Phys. Rev. B}\ }\textbf {\bibinfo {volume} {83}},\ \bibinfo
  {pages} {205101} (\bibinfo {year} {2011})}\BibitemShut {NoStop}%
\bibitem [{\citenamefont {Go}\ \emph {et~al.}(2012)\citenamefont {Go},
  \citenamefont {Witczak-Krempa}, \citenamefont {Jeon}, \citenamefont {Park},\
  and\ \citenamefont {Kim}}]{Go2012}%
  \BibitemOpen
  \bibfield  {author} {\bibinfo {author} {\bibfnamefont {A.}~\bibnamefont
  {Go}}, \bibinfo {author} {\bibfnamefont {W.}~\bibnamefont {Witczak-Krempa}},
  \bibinfo {author} {\bibfnamefont {G.~S.}\ \bibnamefont {Jeon}}, \bibinfo
  {author} {\bibfnamefont {K.}~\bibnamefont {Park}}, \ and\ \bibinfo {author}
  {\bibfnamefont {Y.~B.}\ \bibnamefont {Kim}},\ }\href {\doibase
  10.1103/PhysRevLett.109.066401} {\bibfield  {journal} {\bibinfo  {journal}
  {Phys. Rev. Lett.}\ }\textbf {\bibinfo {volume} {109}},\ \bibinfo {pages}
  {066401} (\bibinfo {year} {2012})}\BibitemShut {NoStop}%
\bibitem [{\citenamefont {Rau}\ \emph {et~al.}(2016)\citenamefont {Rau},
  \citenamefont {Lee},\ and\ \citenamefont {Kee}}]{Rau2016}%
  \BibitemOpen
  \bibfield  {author} {\bibinfo {author} {\bibfnamefont {J.~G.}\ \bibnamefont
  {Rau}}, \bibinfo {author} {\bibfnamefont {E.~K.-H.}\ \bibnamefont {Lee}}, \
  and\ \bibinfo {author} {\bibfnamefont {H.-Y.}\ \bibnamefont {Kee}},\ }\href
  {\doibase 10.1146/annurev-conmatphys-031115-011319} {\bibfield  {journal}
  {\bibinfo  {journal} {Annu. Rev. Condens. Matter Phys.}\ }\textbf {\bibinfo
  {volume} {7}},\ \bibinfo {pages} {195} (\bibinfo {year} {2016})}\BibitemShut
  {NoStop}%
\bibitem [{\citenamefont {Witczak-Krempa}\ \emph {et~al.}(2013)\citenamefont
  {Witczak-Krempa}, \citenamefont {Kim},\ and\ \citenamefont
  {Balents}}]{Witczak-Krempa2013b}%
  \BibitemOpen
  \bibfield  {author} {\bibinfo {author} {\bibfnamefont {W.}~\bibnamefont
  {Witczak-Krempa}}, \bibinfo {author} {\bibfnamefont {Y.~B.}\ \bibnamefont
  {Kim}}, \ and\ \bibinfo {author} {\bibfnamefont {L.}~\bibnamefont
  {Balents}},\ }\href@noop {} {\bibfield  {journal} {\bibinfo  {journal} {Annu.
  Rev. Condens. Matter Phys.}\ }\textbf {\bibinfo {volume} {5}},\ \bibinfo
  {pages} {57} (\bibinfo {year} {2013})}\BibitemShut {NoStop}%
\bibitem [{\citenamefont {Martins}\ \emph {et~al.}(2016)\citenamefont
  {Martins}, \citenamefont {Aichhorn},\ and\ \citenamefont
  {Biermann}}]{Martins2016}%
  \BibitemOpen
  \bibfield  {author} {\bibinfo {author} {\bibfnamefont {C.}~\bibnamefont
  {Martins}}, \bibinfo {author} {\bibfnamefont {M.}~\bibnamefont {Aichhorn}}, \
  and\ \bibinfo {author} {\bibfnamefont {S.}~\bibnamefont {Biermann}},\ }\href
  {\doibase 10.1088/1361-648X/aa648f} {\bibfield  {journal} {\bibinfo
  {journal} {J. Phys.: Condens. Matter}\ }\textbf {\bibinfo {volume} {29}},\
  \bibinfo {pages} {263001} (\bibinfo {year} {2016})}\BibitemShut {NoStop}%
\bibitem [{\citenamefont {Liu}\ and\ \citenamefont
  {Khaliullin}(2018)}]{Liu2018}%
  \BibitemOpen
  \bibfield  {author} {\bibinfo {author} {\bibfnamefont {H.}~\bibnamefont
  {Liu}}\ and\ \bibinfo {author} {\bibfnamefont {G.}~\bibnamefont
  {Khaliullin}},\ }\href {\doibase 10.1103/PhysRevB.97.014407} {\bibfield
  {journal} {\bibinfo  {journal} {Phys. Rev. B}\ }\textbf {\bibinfo {volume}
  {97}},\ \bibinfo {pages} {014407} (\bibinfo {year} {2018})}\BibitemShut
  {NoStop}%
\bibitem [{\citenamefont {Sano}\ \emph {et~al.}(2018)\citenamefont {Sano},
  \citenamefont {Kato},\ and\ \citenamefont {Motome}}]{Sano2018}%
  \BibitemOpen
  \bibfield  {author} {\bibinfo {author} {\bibfnamefont {R.}~\bibnamefont
  {Sano}}, \bibinfo {author} {\bibfnamefont {Y.}~\bibnamefont {Kato}}, \ and\
  \bibinfo {author} {\bibfnamefont {Y.}~\bibnamefont {Motome}},\ }\href
  {\doibase 10.1103/PhysRevB.97.014408} {\bibfield  {journal} {\bibinfo
  {journal} {Phys. Rev. B}\ }\textbf {\bibinfo {volume} {97}},\ \bibinfo
  {pages} {014408} (\bibinfo {year} {2018})}\BibitemShut {NoStop}%
\bibitem [{\citenamefont {Abragam}\ and\ \citenamefont
  {Bleaney}(1970)}]{Abragam1970}%
  \BibitemOpen
  \bibfield  {author} {\bibinfo {author} {\bibfnamefont {A.}~\bibnamefont
  {Abragam}}\ and\ \bibinfo {author} {\bibfnamefont {B.}~\bibnamefont
  {Bleaney}},\ }\href@noop {} {\emph {\bibinfo {title} {Electron Paramagnetic
  Resonance of Transition Ions}}}\ (\bibinfo  {publisher} {Clarendon Press,
  Oxford},\ \bibinfo {year} {1970})\BibitemShut {NoStop}%
\bibitem [{\citenamefont {Perdew}\ \emph {et~al.}(2008)\citenamefont {Perdew},
  \citenamefont {Ruzsinszky}, \citenamefont {Csonka}, \citenamefont {Vydrov},
  \citenamefont {Scuseria}, \citenamefont {Constantin}, \citenamefont {Zhou},\
  and\ \citenamefont {Burke}}]{PBEsol}%
  \BibitemOpen
  \bibfield  {author} {\bibinfo {author} {\bibfnamefont {J.}~\bibnamefont
  {Perdew}}, \bibinfo {author} {\bibfnamefont {A.}~\bibnamefont {Ruzsinszky}},
  \bibinfo {author} {\bibfnamefont {G.}~\bibnamefont {Csonka}}, \bibinfo
  {author} {\bibfnamefont {O.}~\bibnamefont {Vydrov}}, \bibinfo {author}
  {\bibfnamefont {G.}~\bibnamefont {Scuseria}}, \bibinfo {author}
  {\bibfnamefont {L.}~\bibnamefont {Constantin}}, \bibinfo {author}
  {\bibfnamefont {X.}~\bibnamefont {Zhou}}, \ and\ \bibinfo {author}
  {\bibfnamefont {K.}~\bibnamefont {Burke}},\ }\href {\doibase
  10.1103/PhysRevLett.100.136406} {\bibfield  {journal} {\bibinfo  {journal}
  {Phys. Rev. Lett.}\ }\textbf {\bibinfo {volume} {100}},\ \bibinfo {pages}
  {136406} (\bibinfo {year} {2008})}\BibitemShut {NoStop}%
\bibitem [{\citenamefont {http://elk.sourceforge.net}()}]{elk}%
  \BibitemOpen
  \bibfield  {author} {\bibinfo {author} {\bibnamefont
  {http://elk.sourceforge.net}},\ }\href {http://elk.sourceforge,net}
  {}\BibitemShut {NoStop}%
\bibitem [{\citenamefont {Dudarev}\ \emph {et~al.}(1998)\citenamefont
  {Dudarev}, \citenamefont {Botton}, \citenamefont {Savrasov}, \citenamefont
  {Humphreys},\ and\ \citenamefont {Sutton}}]{dudarev}%
  \BibitemOpen
  \bibfield  {author} {\bibinfo {author} {\bibfnamefont {S.~L.}\ \bibnamefont
  {Dudarev}}, \bibinfo {author} {\bibfnamefont {G.~A.}\ \bibnamefont {Botton}},
  \bibinfo {author} {\bibfnamefont {S.~Y.}\ \bibnamefont {Savrasov}}, \bibinfo
  {author} {\bibfnamefont {C.~J.}\ \bibnamefont {Humphreys}}, \ and\ \bibinfo
  {author} {\bibfnamefont {A.~P.}\ \bibnamefont {Sutton}},\ }\href {\doibase
  10.1103/PhysRevB.57.1505} {\bibfield  {journal} {\bibinfo  {journal} {Phys.
  Rev. B}\ }\textbf {\bibinfo {volume} {57}},\ \bibinfo {pages} {1505}
  (\bibinfo {year} {1998})}\BibitemShut {NoStop}%
\bibitem [{\citenamefont {Nirmala}\ \emph {et~al.}(2017)\citenamefont
  {Nirmala}, \citenamefont {Jang}, \citenamefont {Sim}, \citenamefont {Cho},
  \citenamefont {Lee}, \citenamefont {Yang}, \citenamefont {Lee}, \citenamefont
  {Ibberson}, \citenamefont {Kakurai}, \citenamefont {Matsuda}, \citenamefont
  {Cheong},\ and\ \citenamefont {Park}}]{Nirmala}%
  \BibitemOpen
  \bibfield  {author} {\bibinfo {author} {\bibfnamefont {R.}~\bibnamefont
  {Nirmala}}, \bibinfo {author} {\bibfnamefont {K.-H.}\ \bibnamefont {Jang}},
  \bibinfo {author} {\bibfnamefont {H.}~\bibnamefont {Sim}}, \bibinfo {author}
  {\bibfnamefont {H.}~\bibnamefont {Cho}}, \bibinfo {author} {\bibfnamefont
  {J.}~\bibnamefont {Lee}}, \bibinfo {author} {\bibfnamefont {N.-G.}\
  \bibnamefont {Yang}}, \bibinfo {author} {\bibfnamefont {S.}~\bibnamefont
  {Lee}}, \bibinfo {author} {\bibfnamefont {R.~M.}\ \bibnamefont {Ibberson}},
  \bibinfo {author} {\bibfnamefont {K.}~\bibnamefont {Kakurai}}, \bibinfo
  {author} {\bibfnamefont {M.}~\bibnamefont {Matsuda}}, \bibinfo {author}
  {\bibfnamefont {S.-W.}\ \bibnamefont {Cheong}}, \ and\ \bibinfo {author}
  {\bibfnamefont {J.-G.}\ \bibnamefont {Park}},\ }\href {\doibase
  10.1088/1361-648X/aa5c72} {\bibfield  {journal} {\bibinfo  {journal} {J.
  Phys.: Condens. Matter}\ }\textbf {\bibinfo {volume} {29}},\ \bibinfo {pages}
  {13LT01} (\bibinfo {year} {2017})}\BibitemShut {NoStop}%
\bibitem [{\citenamefont {Liu}\ \emph {et~al.}(2008)\citenamefont {Liu},
  \citenamefont {Antonov}, \citenamefont {Jepsen},\ and\ \citenamefont
  {Andersen.}}]{Liu2008}%
  \BibitemOpen
  \bibfield  {author} {\bibinfo {author} {\bibfnamefont {G.-Q.}\ \bibnamefont
  {Liu}}, \bibinfo {author} {\bibfnamefont {V.~N.}\ \bibnamefont {Antonov}},
  \bibinfo {author} {\bibfnamefont {O.}~\bibnamefont {Jepsen}}, \ and\ \bibinfo
  {author} {\bibfnamefont {O.~K.}\ \bibnamefont {Andersen.}},\ }\href {\doibase
  10.1103/PhysRevLett.101.026408} {\bibfield  {journal} {\bibinfo  {journal}
  {Phys. Rev. Lett.}\ }\textbf {\bibinfo {volume} {101}},\ \bibinfo {pages}
  {026408} (\bibinfo {year} {2008})}\BibitemShut {NoStop}%
\bibitem [{\citenamefont {Pesin}\ and\ \citenamefont
  {Balents}(2010)}]{Pesin2010}%
  \BibitemOpen
  \bibfield  {author} {\bibinfo {author} {\bibfnamefont {D.}~\bibnamefont
  {Pesin}}\ and\ \bibinfo {author} {\bibfnamefont {L.}~\bibnamefont
  {Balents}},\ }\href {\doibase 10.1038/nphys1606} {\bibfield  {journal}
  {\bibinfo  {journal} {Nat. Phys.}\ }\textbf {\bibinfo {volume} {6}},\
  \bibinfo {pages} {376} (\bibinfo {year} {2010})}\BibitemShut {NoStop}%
\bibitem [{\citenamefont {Cococcioni}\ and\ \citenamefont
  {de~Gironcoli}(2005)}]{Cococcioni2005}%
  \BibitemOpen
  \bibfield  {author} {\bibinfo {author} {\bibfnamefont {M.}~\bibnamefont
  {Cococcioni}}\ and\ \bibinfo {author} {\bibfnamefont {S.}~\bibnamefont
  {de~Gironcoli}},\ }\href {\doibase 10.1103/PhysRevB.71.035105} {\bibfield
  {journal} {\bibinfo  {journal} {Physical Review B}\ }\textbf {\bibinfo
  {volume} {71}},\ \bibinfo {pages} {035105} (\bibinfo {year}
  {2005})}\BibitemShut {NoStop}%
\bibitem [{\citenamefont {Marzari}\ and\ \citenamefont
  {Vanderbilt}(1997)}]{MLWF}%
  \BibitemOpen
  \bibfield  {author} {\bibinfo {author} {\bibfnamefont {N.}~\bibnamefont
  {Marzari}}\ and\ \bibinfo {author} {\bibfnamefont {D.}~\bibnamefont
  {Vanderbilt}},\ }\href@noop {} {\bibfield  {journal} {\bibinfo  {journal}
  {Phys. Rev. B}\ }\textbf {\bibinfo {volume} {56}},\ \bibinfo {pages} {12847}
  (\bibinfo {year} {1997})}\BibitemShut {NoStop}%
\bibitem [{\citenamefont {Solovyev}\ \emph {et~al.}(1994)\citenamefont
  {Solovyev}, \citenamefont {Dederichs},\ and\ \citenamefont
  {Anisimov}}]{Solovyev1994}%
  \BibitemOpen
  \bibfield  {author} {\bibinfo {author} {\bibfnamefont {I.~V.}\ \bibnamefont
  {Solovyev}}, \bibinfo {author} {\bibfnamefont {P.~H.}\ \bibnamefont
  {Dederichs}}, \ and\ \bibinfo {author} {\bibfnamefont {V.~I.}\ \bibnamefont
  {Anisimov}},\ }\href {\doibase 10.1103/PhysRevB.50.16861} {\bibfield
  {journal} {\bibinfo  {journal} {Phys. Rev. B}\ }\textbf {\bibinfo {volume}
  {50}},\ \bibinfo {pages} {16861} (\bibinfo {year} {1994})}\BibitemShut
  {NoStop}%
\bibitem [{\citenamefont {Park}\ \emph {et~al.}(2012)\citenamefont {Park},
  \citenamefont {Millis},\ and\ \citenamefont {Marianetti}}]{Park2012}%
  \BibitemOpen
  \bibfield  {author} {\bibinfo {author} {\bibfnamefont {H.}~\bibnamefont
  {Park}}, \bibinfo {author} {\bibfnamefont {A.~J.}\ \bibnamefont {Millis}}, \
  and\ \bibinfo {author} {\bibfnamefont {C.~A.}\ \bibnamefont {Marianetti}},\
  }\href {\doibase 10.1103/PhysRevLett.109.156402} {\bibfield  {journal}
  {\bibinfo  {journal} {Phys. Rev. Lett.}\ }\textbf {\bibinfo {volume} {109}},\
  \bibinfo {pages} {156402} (\bibinfo {year} {2012})}\BibitemShut {NoStop}%
\bibitem [{\citenamefont {Haule}\ \emph {et~al.}(2014)\citenamefont {Haule},
  \citenamefont {Birol},\ and\ \citenamefont {Kotliar}}]{Haule2014}%
  \BibitemOpen
  \bibfield  {author} {\bibinfo {author} {\bibfnamefont {K.}~\bibnamefont
  {Haule}}, \bibinfo {author} {\bibfnamefont {T.}~\bibnamefont {Birol}}, \ and\
  \bibinfo {author} {\bibfnamefont {G.}~\bibnamefont {Kotliar}},\ }\href
  {\doibase 10.1103/PhysRevB.90.075136} {\bibfield  {journal} {\bibinfo
  {journal} {Phys. Rev. B}\ }\textbf {\bibinfo {volume} {90}},\ \bibinfo
  {pages} {075136} (\bibinfo {year} {2014})}\BibitemShut {NoStop}%
\bibitem [{\citenamefont {Caffarel}\ and\ \citenamefont
  {Krauth}(1994)}]{dmft_ed}%
  \BibitemOpen
  \bibfield  {author} {\bibinfo {author} {\bibfnamefont {M.}~\bibnamefont
  {Caffarel}}\ and\ \bibinfo {author} {\bibfnamefont {W.}~\bibnamefont
  {Krauth}},\ }\href {\doibase 10.1103/PhysRevLett.72.1545} {\bibfield
  {journal} {\bibinfo  {journal} {Phys. Rev. Lett.}\ }\textbf {\bibinfo
  {volume} {72}},\ \bibinfo {pages} {1545} (\bibinfo {year}
  {1994})}\BibitemShut {NoStop}%
\bibitem [{\citenamefont {Arita}\ \emph {et~al.}(2012)\citenamefont {Arita},
  \citenamefont {Kune\ifmmode~\check{s}\else \v{s}\fi{}}, \citenamefont
  {Kozhevnikov}, \citenamefont {Eguiluz},\ and\ \citenamefont
  {Imada}}]{Arita2012}%
  \BibitemOpen
  \bibfield  {author} {\bibinfo {author} {\bibfnamefont {R.}~\bibnamefont
  {Arita}}, \bibinfo {author} {\bibfnamefont {J.}~\bibnamefont
  {Kune\ifmmode~\check{s}\else \v{s}\fi{}}}, \bibinfo {author} {\bibfnamefont
  {A.~V.}\ \bibnamefont {Kozhevnikov}}, \bibinfo {author} {\bibfnamefont
  {A.~G.}\ \bibnamefont {Eguiluz}}, \ and\ \bibinfo {author} {\bibfnamefont
  {M.}~\bibnamefont {Imada}},\ }\href {\doibase 10.1103/PhysRevLett.108.086403}
  {\bibfield  {journal} {\bibinfo  {journal} {Phys. Rev. Lett.}\ }\textbf
  {\bibinfo {volume} {108}},\ \bibinfo {pages} {086403} (\bibinfo {year}
  {2012})}\BibitemShut {NoStop}%
\bibitem [{\citenamefont {Zhang}\ \emph {et~al.}(2013)\citenamefont {Zhang},
  \citenamefont {Haule},\ and\ \citenamefont {Vanderbilt}}]{Zhang2013}%
  \BibitemOpen
  \bibfield  {author} {\bibinfo {author} {\bibfnamefont {H.}~\bibnamefont
  {Zhang}}, \bibinfo {author} {\bibfnamefont {K.}~\bibnamefont {Haule}}, \ and\
  \bibinfo {author} {\bibfnamefont {D.}~\bibnamefont {Vanderbilt}},\ }\href
  {\doibase 10.1103/PhysRevLett.111.246402} {\bibfield  {journal} {\bibinfo
  {journal} {Phys. Rev. Lett.}\ }\textbf {\bibinfo {volume} {111}},\ \bibinfo
  {pages} {246402} (\bibinfo {year} {2013})}\BibitemShut {NoStop}%
\bibitem [{\citenamefont {P{\"{a}}rschke}\ \emph {et~al.}(2017)\citenamefont
  {P{\"{a}}rschke}, \citenamefont {Wohlfeld}, \citenamefont {Foyevtsova},\ and\
  \citenamefont {van~den Brink}}]{Parschke2017}%
  \BibitemOpen
  \bibfield  {author} {\bibinfo {author} {\bibfnamefont {E.~M.}\ \bibnamefont
  {P{\"{a}}rschke}}, \bibinfo {author} {\bibfnamefont {K.}~\bibnamefont
  {Wohlfeld}}, \bibinfo {author} {\bibfnamefont {K.}~\bibnamefont
  {Foyevtsova}}, \ and\ \bibinfo {author} {\bibfnamefont {J.}~\bibnamefont
  {van~den Brink}},\ }\href {\doibase 10.1038/s41467-017-00818-8} {\bibfield
  {journal} {\bibinfo  {journal} {Nat. Commun.}\ }\textbf {\bibinfo {volume}
  {8}},\ \bibinfo {pages} {686} (\bibinfo {year} {2017})}\BibitemShut {NoStop}%
\bibitem [{\citenamefont {Bergman}\ \emph {et~al.}(2007)\citenamefont
  {Bergman}, \citenamefont {Alicea}, \citenamefont {Gull}, \citenamefont
  {Trebst},\ and\ \citenamefont {Balents}}]{Bergman2007}%
  \BibitemOpen
  \bibfield  {author} {\bibinfo {author} {\bibfnamefont {D.}~\bibnamefont
  {Bergman}}, \bibinfo {author} {\bibfnamefont {J.}~\bibnamefont {Alicea}},
  \bibinfo {author} {\bibfnamefont {E.}~\bibnamefont {Gull}}, \bibinfo {author}
  {\bibfnamefont {S.}~\bibnamefont {Trebst}}, \ and\ \bibinfo {author}
  {\bibfnamefont {L.}~\bibnamefont {Balents}},\ }\href {\doibase
  10.1038/nphys622} {\bibfield  {journal} {\bibinfo  {journal} {Nat. Phys.}\
  }\textbf {\bibinfo {volume} {3}},\ \bibinfo {pages} {487} (\bibinfo {year}
  {2007})}\BibitemShut {NoStop}%
\end{thebibliography}

%

\end{document}